\documentclass[aps,prl,twocolumn,groupedaddress]{revtex4-2}
\usepackage{physics}

\begin{document}


\title{Conformal perturbation theory from open string field theory}


\author{Jaroslav Scheinpflug}
\email[jaroslavscheinpflug at gmail.com]{}
\affiliation{CEICO, Institute of Physics of the Czech Academy of Sciences}
\author{Martin Schnabl}
\email[schnabl.martin at gmail.com]{}
\affiliation{CEICO, Institute of Physics of the Czech Academy of Sciences}


\date{\today}

\begin{abstract}
Conformal boundary conditions in two-dimensional conformal field theories are still mostly an uncharted territory. Even less is known about the relevant boundary deformations that connect them. A natural approach to the problem is via conformal perturbation theory, which however becomes quickly intractable. Using the formalism of open string field theory, we construct the corresponding classical solution, from which we can easily extract the boundary theory data. As a simple illustration we calculate the boundary degeneracy $g$ to next-to-leading order for a generic theory.
\end{abstract}


\maketitle



\newcommand{\re}{\mathop{\rm Re}\nolimits}
\newcommand{\im}{\mathop{\rm im}\nolimits}
\newcommand{\bpz}{\mathop{\rm bpz}\nolimits}
\newcommand{\ad}{\mathop{\rm ad}\nolimits}
\newcommand{\logit}{\mathop{\rm logit}\nolimits}
\newcommand{\asympt}{\mathop{\sim}}
\newcommand{\leftpartial}{\mathop{\!\stackrel{\leftarrow}{\partial}}\nolimits}
\newcommand{\rightpartial}{\mathop{\!\stackrel{\rightarrow}{\partial}}\nolimits}
\newcommand{\leftD}{\mathop{\stackrel{\leftarrow}{D}}\nolimits}
\newcommand{\rightD}{\mathop{\stackrel{\rightarrow}{D}}\nolimits}
\newcommand{\Pexp}{\mathop{\rm Pexp}\nolimits}
\newcommand{\Aexp}{\mathop{\rm \bar{P}exp}\nolimits}
\newcommand{\Ad}{\mathop{\rm Ad}\nolimits}
\def\bra#1{\langle #1 |}
\def\ket#1{|#1 \rangle}
\def\aver#1{\left\langle\, #1 \,\right\rangle}
\def\aaver#1{\left\langle\!\left\langle\, #1 \,\right\rangle\!\right\rangle}
\def\slash#1{\not\!#1}
\def\rdslash{\not\!\partial}
\def\ldslash{\not\!\leftpartial}
\def\ov{\overline}
\def\res#1{\oint\! \frac{d#1}{2\pi i} \,}
\def\half{\frac{1}{2}}
\newcommand{\VEV}[1]{\left\langle #1\right\rangle}
\newcommand{\inomega}[1]{\sqrt{\Omega}#1\sqrt{\Omega}}

\let\eps = \epsilon
\def\beps  {\bar\epsilon}
\def \be {\begin{equation}}
\def \ee {\end{equation}}
\def \bea {\begin{eqnarray}}
\def \eea {\end{eqnarray}}
\def \bdm {\begin{displaymath}}
\def \edm {\end{displaymath}}
\def \ot{\otimes}

\def \nn {{\mathbb N}}
\def \zz {{\mathbb Z}}
\def \cc {{\mathbb C}}
\def \rr {{\mathbb R}}
\def \TT {{\mathbb T}}
\def \tr{{\rm tr}}
\def \ii {{\cal I}}
\def \ll {{\cal L}}
\def \dd {{\cal D}}
\def \kk {{\cal K}}
\def \hh {{\cal H}}
\def \aa {{\cal A}}
\def \bb {{\cal B}}
\def \mm {{\cal M}}
\def \oo {{\cal O}}
\def \pp {{\cal P}}
\def \nnn {{\cal N}}
\def \eee {{\cal E}}
\def \ccc {{\cal C}}
\def \sss {{\cal S}}
\def \uu {{\cal U}}
\def \nc {noncommutative }
\def \ncg {noncommutative geometry }
\def \sf  {string field }
\def \sft {string field theory }
\def \da {\dagger}
\def\del {\partial}
\def\0{\nonumber}
\def \AA {{\mathbb A}}
\def \QQ {{\mathbb Q}}
\def \XX {{\mathbb X}}
\def \lL {{\mathbb L}}
\def \VV {{\mathbb V}}
\def \BB {{\mathbb B}}
\def \BCFT{{\rm BCFT}}
\def\SFT{{\rm SFT}}
\def\f{{\bar f}}
\def\F{{\mathcal F}}
\def\G{{\mathcal G}}
\def\S{{\mathcal S}}
\def\P{{\mathcal P}}
\newcommand{\cn}{\mathop{\rm cn}\nolimits}
\def\curlypropag{{-\bar{P}\frac{\mathcal{B}_0}{\mathcal{L}_0}}}
\def\curlypropagnm{{\bar{P}\frac{\mathcal{B}_0}{\mathcal{L}_0}}}
\newcommand{\normord}[1]{:\mathrel{#1}:}
\newcommand{\wedgeprefac}[1]{U_{#1}^{*}U_{#1}}
\newcommand{\argpare}[1]{\left(#1\right)}
\newcommand{\Ellwood}{\mathop{\rm Tr_\mathcal{V}}\nolimits}
\def\twopt{}
\def\threept{C_{VVV}}
\def\threeptW{C_{VVV'}}

\def \Pbar {{\bar P}}

\section{Introduction and Conclusion}

Study of consistent boundary conditions in two-dimensional conformal field theory (CFT) is a long and fascinating subject with many applications. There is a number of stringent constraints on the spectrum of boundary operators and operator product expansion (OPE) coefficients \cite{Cardy:1986gw, Cardy:1989ir, Lewellen:1991tb}. Flurry of work towards the end of the last millennium culminated in a fairly complete characterization of boundary conditions in rational conformal field theories \cite{Cardy:1989ir, Behrend:1999bn}. Little is known, however, for the non-rational case. Except in few examples \cite{Affleck:2000ws, Schnabl:2019oom} where for special points in the moduli space rational structure may emerge, one has to resort either to perturbative or numerical approaches. Numerically, one can use the truncated conformal space approach \cite{Dorey:1997yg, Runkel} or level truncation in open string field theory (OSFT) \cite{Kudrna:2014rya, Kudrna:2021rzd}. Perturbatively, one can start with a given consistent conformal boundary condition and perturb it by a weakly relevant operator with dimension $h$ close to 1. Such flows were first considered by Affleck and Ludwig \cite{Affleck:1991tk} following analogous calculations by Zamolodchikov \cite{Zamolodchikov:1987ti} for the bulk perturbations.
They have introduced a new universal quantity $g$ characterizing conformal boundary conditions, playing a role of noninteger ground-state degeneracy. It appeared as an intensive part of the free energy on a cylinder and can be interpreted as the value of the disk partition function.
For a nearby fixed point they showed
\be
\frac{\Delta g}{g} = - \frac{\pi^2}{3} \frac{y^3}{C_{VVV}^2} + O(y^4), \qquad y \equiv 1-h,
\label{CPTg}
\ee
where $C_{VVV}$ is a three-point structure constant of $V$,
 which led them to conjecture that the corresponding $g$-function decreases along the RG flow which was later proved \cite{Friedan:2003yc}. These flows have been applied to minimal models \cite{Recknagel:2000ri} and further studied in \cite{Schmidt-Colinet:2009pga}

Open string field theory is a consistent theory of open strings ending on D-branes. The string degrees of freedom describe collective degrees of freedom for these nonperturbative objects, such as its position and shape in the spacetime, or fields living on them. To formulate open string field theory one needs a reference D-brane. The vector space of quantum open strings ending on the D-brane is given by the Hilbert space of boundary conformal field theory. The classical field $\Psi$ of open string field theory is an element from this space. Different backgrounds correspond to different classical solutions of the equation of motion $Q\Psi + \Psi*\Psi =0$. The value of the Witten's action
\be
S = - \frac{1}{2} \aver{\Psi, Q\Psi} - \frac{1}{3} \aver{\Psi, \Psi*\Psi}
\label{Action}
\ee
for a given classical solution has been conjectured by Sen \cite{Sen:1999xm} and later proved by others \cite{Schnabl:2005gv,Erler:2014eqa} (see \cite{Erler:2019vhl} for a recent review) to correspond to a change in the given boundary degeneracy between the new and old boundary conditions
\be
S = - \frac{\Delta g}{2\pi^2}.
\ee
In this work we show that open string field theory can be highly efficient in circumventing the notorious difficulties in higher order conformal perturbation theory by providing a manifestly finite all-order scheme. We construct perturbatively the solution corresponding to a nearby fixed point and calculate the change in the $g$-function
\be
\frac{\Delta g}{g_0} = -\frac{\pi^2}{3} \left( \frac{y^3}{C_{VVV}^2}  + \frac{3\mathcal{\tilde A}}{C_{VVV}^4} y^4 + O(y^5)\right),
\label{nextToLeadingG}
\ee
where $g_0$ is the $g$-function of the undeformed theory and
\bea
\mathcal{\tilde A} &=& \int_0^{1/2} d\xi \biggr[\frac{1}{g_0}\aver{V| V(1) V(\xi) |V} \nonumber\\
&&- \sum_{V'} C_{VVV'}^2 \left(\frac{1}{\xi^{2-h'}} + \frac{1}{(1-\xi)^{2-h'}} \right)\biggr]
\nonumber\\&& +\sum_{V' \ne V} C_{VVV'}^2 \left(\frac{1}{h'-1} + \frac{1}{2} \delta_{h'=0} \right).
\eea
The first sum over the relevant fields in the OPE of two $V$'s renders the integral fully convergent. The subtraction of powers of $1-\xi$ has been introduced for mere convenience. The contribution of the relevant fields is uniquely fixed by our OSFT calculations and is accounted for by the second sum.
We have applied (\ref{nextToLeadingG}) to RG flows of $c=2$ free bosons and obtained a high precision match with the unpublished numerics of \cite{Exotic}.

The results of \cite{KMS} also allow us to compute the leading order correction to the coefficient of a bulk primary $\phi$ in the boundary state
\be
\begin{aligned}
    \frac{\Delta B_{\phi}}{g_0} &= - 2\pi \frac{B_{\phi V} \twopt}{\threept} y + O(y^2),
\end{aligned}
\ee
where $B_{\phi V}$ is a bulk-boundary structure constant. The next-to-leading correction should be straightforward to obtain with the methods of this paper.

\section{General setup}

A general BCFT of central charge $c$ can be viewed as a part of open string background, if accompanied by an auxiliary BCFT with central charge $26-c$. The Hilbert space of the open string contains also the sector of $b,c$ ghosts with $c=-26$, so that the total central charge is zero. Classical solutions of OSFT equation of motion $Q\Psi+\Psi*\Psi=0$, where $Q$ is the BRST charge and $*$ is the Witten's star product, correspond to changes in the background.

A family of classical solutions can be constructed perturbatively, by writing $\Psi = R + X$, where $R = \sum_i \lambda_i V^{i} c_1 \ket{0}$ is a finite sum of states corresponding to the perturbing nearly-marginal operators $V^i$ and $X$ stands for the remainder.
Using $\lbrace Q, b_0 \rbrace =L_0$ and imposing the Siegel gauge condition $b_0 \Psi =0$ the equation of motion implies
\be
L_0 \Psi + b_0 (\Psi*\Psi) = 0.
\ee
This can be rewritten as
\bea
\label{Req}
R + \frac{b_0}{L_0} P (R+X)^2 &=& 0, \\
X + \frac{b_0}{L_0} \Pbar (R+X)^2 &=& 0,
\label{Xeq}
\eea
where from now on we mostly omit the star product symbol. Here $P$ is a projector onto the vector space spanned by $V^{i} c_1 \ket{0}$ and $V^{i} c_1 c_0 \ket{0}$, and $\Pbar = 1 - P$. The second equation can be solved perturbatively:
\be\label{Xsol}
X =  H R^2 + H \left\{ R, H R^2 \right\} + \cdots, \quad  H \equiv - \frac{b_0}{L_0} \bar P.
\ee
The right hand side of (\ref{Xsol}) corresponds to all possible binary diagrams, where each vertex denotes star multiplication, each internal line an action of $H$ and each external line $R$. For a given number $n$ of external lines (number of $R$'s), there is $C_{n-1}$ number of such terms where $C_n = 1,2,5,\ldots$ are the Catalan numbers. The numbers would appear as coefficients of the Taylor series of the solution to (\ref{Xeq}) if one treated the string fields $R$ and $X$ and the operator $H$ as plain numbers.

The Catalan numbers grow asymptotically only as $C_n \sim \frac{4^n}{n^{3/2} \sqrt{\pi}}$, so we see that the series (\ref{Xsol}) should be convergent, for sufficiently small $R$ (in some convenient norm) assuming that the action of $H$ is bounded from above. This should be the case when the nearest marginal states $V^{i} c\partial c \ket{0}$ are projected out from the ghost number two Hilbert space on which $H$ acts in (\ref{Xsol}).
The key is that the projector $\Pbar$ projects out the nearly marginal states for which $1/L_0$ would have the biggest eigenvalue. From this rough argument it is clear that the coefficients of $R$ and hence also $\Psi$ should be parametrically smaller than $1-\tilde h$, where $\tilde h$ is the dimension of the nearest marginal operator surviving the $\Pbar$ projection.
For theories with marginal operators, we would need either to guarantee that those operators are not excited in our solution, or alternatively, change the definition of the projector $\Pbar$ so that it would project out such states as well.

Plugging the solution (\ref{Xsol}) into (\ref{Req}) we find a system of finite number of equations (generally non-polynomial if we do not truncate to finite order in $R$). This has the interpretation of the fixed point equations for the relevant couplings $\lambda_i$.

While it is immediately obvious that the full equations for $R$ are obeyed
\be\label{QReq}
Q R + P (R+X)^2 = 0
\ee
as a consequence of (\ref{Req}) and the fact $\im(P) \subset \ker(Q)$  for ghost number two states, it is not a priori obvious that
\be
Q X + \bar P (R+X)^2 = 0.
\ee
In fact, in general it is not true, that any given solution of the gauge-fixed equations of motion obeys the full equation of motion \cite{Kudrna:2018mxa}. In the case of perturbatively constructed solutions, however, one can prove it \cite{Erbin:2020eyc}. To show this, note that from (\ref{Xeq}) and (\ref{QReq})
\be
Q \Psi + \Psi^2 = -H Q (\Psi^2) = H [ H Q (\Psi^2) , \Psi] = \cdots,
\ee
where in the last step we used the derivation property of $Q$, the already proven part of the equation of motion, and dropped $[\Psi^2,\Psi]=0$. Now assuming that $\Psi$ is parametrically small enough in some convenient norm, and that repeated action of $H$ does not give rise to divergent factors, one can argue that the right hand side vanishes. Unlike in the discussion below (\ref{Xsol}), here $H$ acts always on
$Q$-exact states at ghost number 3. The state of the form $c\partial c \partial^2 c$ with lowest possible value of $L_0$ cannot appear since it is not $Q$-exact. States of the form $c\partial c \partial^2 c V$ for $V$ of low conformal weight (most relevant) are $Q$-exact, so it is important that there is a gap between the identity operator with $h=0$ and the next most relevant operator in the matter CFT vector space. Practically the bounds are even better, since states of the form are excluded by twist symmetry which is often imposed for useful solutions.

\section{Conformal perturbation theory}

We start by writing our Siegel-gauge solution for the perturbative flow triggered by the single nearly marginal field $V$ with $h \sim 1$ as $\Psi = \sum_{n=1}^{\infty} \Psi_n$
where $\Psi_1 \equiv R = \lambda cV\ket{0}$ and
\be
\Psi_2 = H R^2,\quad  \Psi_3 = H \lbrace R, H R^2 \rbrace, \quad \ldots.
\ee
By construction $\Psi_n = O(\lambda^n)$. For the perturbing field $V$ we assume the fusion rule 
$V \times V = 1 + V + \cdots$,
where the dots stand for possible other fields which are not nearly marginal.
We normalize our upper-half-plane correlators as
\bea
\nonumber
\langle V(z_1)V(z_2)\rangle &=& g_0 \twopt z_{12}^{-2h}, \\
\label{3pt}
\langle V(z_1)V(z_2)V(z_3) \rangle &=& g_0 \twopt \threept z_{12}^{-h}z_{13}^{-h}z_{23}^{-h},
\eea
where here and in the rest of the paper we write $z_{ij} \equiv \abs{z_i-z_j}$ for the matter fields and $z_{ij} \equiv z_i - z_j$ in the analogous formulas for ghosts.
To make contact with the usual conformal perturbation theory we would like to compute observables in perturbation series with an expansion parameter $y =1-h$. To do this we first need to compute the SFT coupling $\lambda$ as a function of $y$.

\subsection{Calculation of the SFT coupling}

Contracting the (\ref{QReq}) for $R \equiv \Psi_1 = \lambda c V \ket{0}$ with $cV$ and using $Q\Psi_1 = y \lambda c \del c V\ket{0}$ we find
\be
y\lambda \langle  c V, c \del c V \rangle + \lambda^2 \langle cV, cV * cV\rangle  - 2\lambda^3 \mathcal{A} + O(\lambda^4) = 0,
\label{fPointEq}
\ee
where we introduced the open string off-shell four-point amplitude
\be
\mathcal{A} \equiv \langle c V * cV,  \frac{b_0}{L_0} \bar{P} (cV * cV)\rangle.
\ee
Using the correlators 
(\ref{3pt}) and $\langle c(z_1)c(z_2)c(z_3)\rangle = z_{12} z_{13} z_{23}$
we find the SFT coupling 
\be
    \lambda = \frac{y}{C_{VVV}} + \frac{1}{C_{VVV}}\left(\frac{2\mathcal{A}}{g_0 \twopt\threept^2} - \ln K^3\right)y^2 + O(y^3),
\label{coupling}
\ee
where $K \equiv \frac{3 \sqrt{3}}{4}$ is the ubiquitous string field theory constant appearing in the start products such as $\langle cV, cV * cV\rangle = K^{3-3h} g_0 C_{VVV}$.  To leading order this is precisely the result obtained by Affleck and Ludwig \cite{Affleck:1991tk} modulo a sign convention so that at leading order the SFT and CFT couplings are equal in magnitude. Next we move on to the calculation of gauge invariant observables.
\subsection{Calculation of gauge invariant observables}
In this section we evaluate the perturbative shift in the $g$-function triggered by a deformation by $V$ to next-to-leading order in $y$. It turns out that the OSFT action is very efficient in accomplishing this. We also calculate the corresponding leading order shift in the boundary state coefficients by calculating the Ellwood invariant and invoking the Kudrna-Maccaferri-Schnabl (KMS) correspondence \cite{KMS}.

\subsubsection{Leading order calculation of $g$}
The on-shell value of the action (\ref{Action}) of a solution $\Psi$ can be written as $S = -\frac{1}{6}\langle \Psi, Q\Psi\rangle$
so that the corresponding shift in the $g$-function is
\be
\Delta g = -2\pi^2 S = \frac{\pi^2}{3} \langle \Psi, Q\Psi\rangle,
\ee
which using $\langle \Psi_1, Q\Psi_1\rangle = -g_0 y \lambda^2$ and (\ref{coupling})
gives
\be
\frac{\Delta g}{g_0} = - \frac{\pi^2}{3} \frac{y^3}{C_{VVV}^2} + O(y^4),
\ee
matching the leading order result of \cite{Affleck:1991tk}.

\subsubsection{Next-to-leading order calculation of $g$}
In the next-to-leading order we have
\be
\begin{aligned}
\Delta g =& \frac{\pi^2}{3} \biggr(\langle \Psi_1, Q\Psi_1\rangle + \langle \Psi_2,Q\Psi_2\rangle\biggr) + O(y^5),
\end{aligned}
\ee
where we have dropped the off-diagonal terms by the BPZ-orthogonality of $P$. We can also write
\be
\langle \Psi_2, Q\Psi_2\rangle = - \langle \Psi_2, \bar{P} \Psi_1 * \Psi_1\rangle = - \langle \Psi_2, \Psi_1 * \Psi_1\rangle =  \lambda^4 \mathcal{A}
\label{druheA}
\ee
so that we find
\be
\frac{\Delta g}{g_0} = -\frac{\pi^2}{3}\biggr(\frac{y^3\twopt}{\threept^2} + \biggr(\frac{3\mathcal{A}}{g_0\threept^4} -  \frac{\ln K^6 \twopt}{\threept^2} \biggr)y^4\biggr) + O(y^5).
\ee

It remains to calculate the open string four-point amplitude $\mathcal{A}$. Using some standard OSFT technology
\cite{Rastelli:2000iu, Schnabl:2002gg, Berkovits:2003ny} we can write
\be
\mathcal{A} = \langle cV\argpare{-\sqrt{3}} cV\argpare{\sqrt{3}} \,Z\, cV\argpare{\frac{1}{\sqrt{3}}}cV\argpare{-\frac{1}{\sqrt{3}}}\rangle,
\ee
where we define the object $Z =-U_3 H U_3^{*}$ 
with $U's$ some specific exponentials of sums of total Virasoro generators, which commute with $Q$ but not with $b_0/L_0$.
Then we use a variant of the trick of \cite{Berkovits:2003ny} by inserting the Hodge-Kodaira decomposition
\be
1 = \acomm{Q}{\frac{b_0}{L_0}\bar{P}} + P
\ee
on both sides of $Z$ and using $H U_3 P = P U_3^* H = 0$ to obtain
\be
\begin{aligned}
Z = &- H U_3 U_3^* \frac{1+P}{2} - \frac{1+P}{2} U_3 U_3^* H + P Z P \\
&- \comm{Q}{H Z \frac{1+P}{2} - \frac{1+P}{2}Z H}
\end{aligned}.
\label{Zko}
\ee
As $cV$ is $Q$-closed up to $O(y)$ terms, we can ignore the last term of (\ref{Zko}). The contribution of the $PZP$ term in (\ref{Zko}) can be readily simplified since
\be
P cV\argpare{\frac{1}{\sqrt{3}}}cV\argpare{-\frac{1}{\sqrt{3}}} = -\threept \argpare{\frac{2}{\sqrt{3}}}^{1-h} \! c\del c V (0),
\label{PcVcV}
\ee
which follows from the OPE
\be
cV(x)cV(-x) \sim -\sum_{V'}  \threeptW \frac{c\del c V'(0)}{(2x)^{2h-h'-1}}
\label{OPE}
\ee
so that its contribution to $\mathcal{A}$ is
\be
\threept^2 \argpare{\frac{4}{3}}^y \bra{c\del c V} U_3 \frac{b_0}{L_0}\bar{P} U_3^* \ket{c\del cV}.
\label{Acontrib3}
\ee
Writing out $\bar{P} = 1 - P$ with $c \partial c V = \frac{1}{y} Q(c V)$ and noting that by the explicit Virasoro formulas \cite{Rastelli:2000iu, Schnabl:2002gg} for $U_n^*$ we have $
PU_3^*\ket{c\del c V} = \argpare{\frac{2}{3}}^{h-1}\ket{c\del c V}$
and by commuting the $U$s as $U_3 U_3^{*} = U_{\frac{8}{3}}^* U_{\frac{8}{3}}$ \cite{Schnabl:2002gg} we get for (\ref{Acontrib3})
\be
\begin{aligned}
& -g_0\twopt \threept^2 \argpare{\frac{4}{3}}^y\frac{1}{y} \argpare{\argpare{\frac{3}{4}}^{-2y}-\argpare{\frac{2}{3}}^{-2y}} \\ &= g_0\twopt\threept^2 \ln \frac{81}{64} + O(y).
\end{aligned}
\label{gammaCalc}
\ee
In an unpublished work \cite{Gamma} Maccaferri et al. argued on general grounds that there exists a geometric constant $\gamma$ independent of the underlying matter CFT such that
\be
\bra{c\del c V} U_3\left(\frac{b_0}{L_0}\bar{P}\right)U_3^{*} \ket{c\del c V} = \gamma \bra{c \del c V} b_0 \ket{ c\del c V}.
\ee
Our result $\gamma = \ln \frac{81}{64}$ agrees with their level truncation calculations to seven decimal places. This computation was made possible by the fact that our field was not marginal making intermediate expressions such as $\frac{1}{h-1}$ well-defined.

We continue with the contribution coming from the first two terms of (\ref{Zko}), which give the same contribution because of BPZ symmetry, and in doing so after using $U_3 U_3^* = U_{\frac{8}{3}}^* U_{\frac{8}{3}}$, we encounter
\be
\begin{aligned}
&U_{\frac{8}{3}}\frac{b_0}{L_0}\bar{P} cV\argpare{\frac{1}{\sqrt{3}}}cV\argpare{-\frac{1}{\sqrt{3}} }\ket{0} = \\
& U_{\frac{8}{3}} \frac{b_0}{L_0} \bar{P} \biggr[cV\argpare{\frac{1}{\sqrt{3}}}cV\argpare{-\frac{1}{\sqrt{3}}} \\
&\pm \sum_{V' \neq V}  \threeptW \argpare{\frac{2}{\sqrt{3}}}^{1+h' - 2h}c\del c V'(0)\biggr]\ket{0},
\end{aligned}
\ee
where we subtracted and added the singular part of the OPE  (\ref{OPE}) in the image of $\bar{P}$. This allows us to use the Schwinger parametrization $
\frac{1}{L_0} = \int_{0}^{1}\,\frac{ds}{s} s^{L_0}$
on the original term with the subtraction (plus term). On the minus term we can invert $L_0$ explicitly since each term in the sum is an $L_0$ eigenstate. Doing so we obtain
\be
\begin{aligned}
&\int_{0}^{1}\, ds \biggr[\argpare{\dv{\mu}{s}}^{2h-1}\frac{s^{2h-2}}{\sqrt{3}}\argpare{c\argpare{\mu}+c\argpare{-\mu}}V\argpare{\mu}V\argpare{-\mu} \\
&- \sum_{V' \neq V}  \threeptW \argpare{\frac{2}{\sqrt{3}}}^{3 -h' - 2h} \biggr(\frac{1}{s^{2-h'}} + \frac{1}{1-h'}\biggr) cV'(0) \\
&  - \threept \argpare{\frac{2}{\sqrt{3}}}^{3-3h} \frac{1}{s^{2-h}}cV(0) \biggr]\ket{0},
\end{aligned}
\ee
where $\mu(s) = \tan \frac{3}{4} \arctan \frac{s}{\sqrt{3}}$ is present since $U_{\frac{8}{3}}$ implements the conformal transformation $z \rightarrow \tan \frac{3}{4} \arctan z$.

The subterm in the contribution coming from the first two terms of (\ref{Zko}) containing $P$ can then be calculated as
\be
\begin{aligned}
    &-\threept \argpare{\frac{8}{3\sqrt{3}}}^{1-h} \bra{c \del c V} \\ & \int_{0}^{1}\,\hspace{-0.25cm}ds \biggr[\argpare{\dv{\mu}{s}}^{2h-1}\frac{s^{2h-2}}{\sqrt{3}}\argpare{c\argpare{\mu}+c\argpare{-\mu}}V\argpare{\mu}V\argpare{-\mu} \\
    &- \sum_{V' \neq V}  \threeptW \argpare{\frac{2}{\sqrt{3}}}^{3 -h' - 2h} \biggr(\frac{1}{s^{2-h'}} + \frac{1}{1-h'}\biggr) cV'(0) \\
    &  - \threept \argpare{\frac{2}{\sqrt{3}}}^{3-3h} \frac{1}{s^{2-h}}cV(0) \biggr]\ket{0} \\ &= g_0\twopt \threept^2 \ln \frac{4a}{\sqrt{3}} + O(y),
\end{aligned}
\label{subtermP}
\ee
where $a \equiv \mu(1) = \tan \frac{\pi}{8} = \sqrt{2}-1$.

The remaining term to calculate is then up to $O(y)$ terms
\begin{widetext}
\be
\begin{aligned}
    &\langle cV\argpare{\sqrt{3}}cV\argpare{-\sqrt{3}}U_\frac{8}{3}^* \int_{0}^{1}\,\hspace{-0.25cm}ds \biggr[\argpare{\dv{\mu}{s}}^{2h-1}\frac{s^{2h-2}}{\sqrt{3}}\argpare{c\argpare{\mu}+c\argpare{-\mu}}V\argpare{\mu}V\argpare{-\mu} \\
     & - \sum_{V' \neq V}  \threeptW \argpare{\frac{2}{\sqrt{3}}}^{3-h' - 2h} \biggr(\frac{1}{s^{2-h'}} + \frac{1}{1-h'}\biggr) cV'(0)
       - \threept \argpare{\frac{2}{\sqrt{3}}}^{3-3h} \frac{1}{s^{2-h}}cV(0) \biggr]\rangle
      \\ &= \int_{0}^{1}\,\hspace{-0.25cm}ds \biggr[\frac{4}{\sqrt{3} a}\! \argpare{\frac{1}{a^2}-\mu^2} \dv{\mu}{s} \langle V\!\argpare{-\frac{1}{a}}V\!\argpare{\frac{1}{a}} V\argpare{\mu} V\argpare{-\mu}\rangle - g_0\sum_{V' \neq V}  \twopt \threeptW^2 \argpare{\sqrt{3} a}^{h' - 1}\biggr(\frac{1}{s^{2-h'}} + \frac{1}{1-h'}\biggr)  - \twopt \threept^2 \frac{g_0}{s}\biggr].
    \label{As}
\end{aligned}
\ee
\end{widetext}
It turns out that we can exploit the generic form of the 4-pt function \cite{DiFrancesco}
\be
\langle V(z_1)V(z_2)V(z_3)V(z_4)\rangle =
\frac{G(\xi)}{\argpare{z_{12}z_{13}z_{14}z_{23}z_{24}z_{34}}^{\frac{2}{3}h}},
\ee
where $\xi$ is the cross-ratio
\be
\xi = \frac{z_{12}z_{34}}{z_{13}z_{24}}
\ee
and $G(\xi)$ is an invariant part of the 4-pt function, to dramatically simplify
\be
\begin{aligned}
    &\int_{0}^{1}\,\hspace{-0.25cm}ds \biggr[\frac{4}{\sqrt{3} a} \argpare{\frac{1}{a^2}-\mu^2} \dv{\mu}{s} \langle V\argpare{-\frac{1}{a}}V\argpare{\frac{1}{a}} V\argpare{\mu} V\argpare{-\mu}\rangle\biggr] \\ &= \int_{0}^{\frac{1}{2}}\, d\xi \bra{V} V(1)V(\xi) \ket{V}.
\end{aligned}
\ee
To make this expression convergent, we must not forget the subtractions in (\ref{As}).
We could now transform the subtractions via inverting $\xi(s)$, where the cross-ratio depends on $s$ as follows
\be
\xi(s) = 4 a\frac{\mu(s)}{\left(1+a \mu(s)\right)^2},
\ee
but this would lead to unwieldy expressions. Equivalently, we employ a near-zero cutoff $\tilde\epsilon$ of the manifestly finite integral (\ref{As}) to separate the subtractions from the 4-pt function into two integrals and in the end we take $\epsilon \equiv \xi(\tilde\epsilon)$ to zero.
Integrating the subtractions in (\ref{As}) we get
\be
\begin{aligned}
&\sum_{V'\neq V} g_0 \twopt \threeptW^2 \argpare{\sqrt{3} a}^{h' - 1} \frac{\tilde\epsilon^{h'-1}}{h'-1} + g_0 \twopt \threept^2 \ln \tilde\epsilon \\
&= \sum_{V'\neq V} g_0 \twopt \threeptW^2\biggr(\frac{\epsilon^{h'-1}}{h' - 1} + \frac{1}{2} \delta_{h'=0}\biggr) \\ & \quad + g_0 \twopt \threept^2 \left(\ln \epsilon + \ln \frac{\sqrt{2}+1}{\sqrt{3}}\right)  + O(\epsilon),
\end{aligned}
\label{regsub}
\ee
where we used
\be
\tilde\epsilon(\epsilon) = \sqrt{1+\frac{2 \sqrt{2}}{3}} \, \epsilon +\sqrt{\frac{1}{4}+\frac{1}{3 \sqrt{2}}} \, \epsilon ^2 + O(\epsilon^3).
\ee
The integration of the 4-pt function over the cross-ratio $\xi$ can be simply rewritten as
\be
\begin{aligned}
&\int_{\epsilon}^{\frac{1}{2}}\, d\xi \biggr(\bra{V} V(1)V(\xi) \ket{V} - g_0\sum_{V'} \twopt\threeptW^2\frac{1}{\xi^{2-h'}}\biggr) \\ &+ g_0\sum_{V'\neq V} \twopt \threeptW^2 \argpare{\frac{2^{1-h'}}{h'-1}-\frac{\epsilon ^{h'-1}}{h'-1}} \\ &+ g_0\twopt\threept^2 \argpare{\ln 2 - \ln \epsilon}.
\end{aligned}
\label{regint}
\ee
We see that the divergent terms cancel each other out exactly, so that we can take $\epsilon \rightarrow 0$. Putting all contributions (\ref{gammaCalc}), (\ref{subtermP}), (\ref{regsub}), and (\ref{regint}) together we obtain
\be
\begin{aligned}
&\mathcal{A} = \int_{0}^{\frac{1}{2}}\, d\xi \biggr[\bra{V} V(1)V(\xi) \ket{V} \\ &- g_0\sum_{V'} \twopt\threeptW^2\biggr(\frac{1}{\xi^{2-h'}} + \frac{1}{(1-\xi)^{2-h'}}\biggr)\biggr] \\ &+ g_0\sum_{V'\neq V} \twopt \threeptW^2 \biggr(\frac{1}{h'-1}+\frac{1}{2}\delta_{h'=0}\biggr) \\ &+ g_0\twopt\threept^2\argpare{\ln \frac{\sqrt{2}+1}{\sqrt{3}} + \ln \frac{81}{64} + \ln \frac{4a}{\sqrt{3}}} + O(y),
\end{aligned}
\ee
where the $1-\xi$ subtractions were added for convenience. The last term in the bracket simplifies to $2\ln K$. As a trivial illustration of this formula
we consider the exactly marginal free boson $V = \del X$ for which
\be
g_0^{-1} \bra{\del X} \del X\argpare{1} \del X\argpare{\xi} \ket{\del X}=  1+\frac{1}{(1-\xi)^2}+\frac{1}{\xi^2},
\ee
so that the four-point amplitude vanishes
\be
g_0^{-1} \mathcal{A} = \left( \int_{0}^{\frac{1}{2}}\, d\xi \right) -1 + \frac{1}{2} = 0,
\ee
as it should, since (\ref{fPointEq}) is then automatically satisfied for all $\lambda$ to the next-to-leading order.

The final result for $\Delta g$ then simplifies to
\be
\frac{\Delta g}{g_0} = - \frac{\pi^2}{3} \biggr(\frac{y^3 \twopt}{\threept^2} + 3 \frac{\mathcal{\tilde A}}{\threept^4}y^4\biggr) + O(y^5),
\label{gNextToLeading}
\ee
where
\be
\begin{aligned}
&\mathcal{\tilde A} \equiv g_0^{-1}\mathcal{A} - 2 \threept^2 \ln K \\
& = \int_{0}^{\frac{1}{2}}\, d\xi \biggr[\frac{1}{g_0}\bra{V} V(1)V(\xi) \ket{V} \\ &- \sum_{V'} \twopt\threeptW^2\biggr(\frac{1}{\xi^{2-h'}} + \frac{1}{(1-\xi)^{2-h'}}\biggr)\biggr] \\ &+ \sum_{V'\neq V} \twopt \threeptW^2 \biggr(\frac{1}{h'-1}+\frac{1}{2}\delta_{h'=0}\biggr).
\end{aligned}
\label{ATilde}
\ee
Note that the constant $K$ disappeared from the final expression for the physical observable $g$, but
is still present in the expression for the SFT coupling
\be\label{lambdaA}
\lambda = \frac{y}{C_{VVV}} + \frac{1}{C_{VVV}}\left(\frac{2\mathcal{\tilde A}}{\twopt\threept^2} + \ln K\right)y^2 + O(y^3).
\ee
This is possible since the coupling constant of the perturbing operator is in general scheme dependent, and OSFT provides one particular consistent scheme.


\subsubsection{Leading order calculation of boundary state coefficients}
By the KMS correspondence \cite{KMS}, the shift in the boundary state coefficient of the bulk primary $\phi$ is
\be
\Delta B_{\phi} = 2 \pi i \bra{I} \tilde\phi(i) \ket{\Psi},
\ee
where $\tilde\phi$ is $\phi$ dressed with ghosts and auxiliary CFT factors that make it weight 1 and $\bra{I} = \bra{0}U_f$ implements the conformal transformation $f(z) = \frac{2 z}{1-z^2}$. To leading order we can write
\be
\begin{aligned}
    \Delta B_{\phi} &= 2\pi i\lambda\bra{0} U_f\tilde\phi(i)\ket{cV} + O(y^2) \\
                    &= 2\pi i \lambda 2 i 2^{\Delta_{\phi}-2} 2^{h-1} \langle \phi(i) V(0) \rangle  + O(y^2)\\
                    &= - 2\pi g_0 \frac{B_{\phi V} \twopt}{\threept} y + O(y^2),
\end{aligned}
\label{BoundaryLeading}
\ee
where $\Delta_{\phi}$ is the weight of $\phi$ and in the derivation we used the generic form of the bulk-boundary correlator
\be
\langle \phi(x+i y) V(r) \rangle = g_0 \frac{B_{\phi V}\twopt  }{(2y)^{\Delta_{\phi} - h} \left((x-r)^2+y^2\right)^h}.
\ee
For the case of $\phi$ being the identity operator, we expect (\ref{BoundaryLeading}) to calculate $\Delta g$ to $O(y)$ and indeed it gives the expected result of zero by the virtue of $B_{1V} = 0$. Our result (\ref{lambdaA}) for the coupling is an important ingredient together with the open-open-closed amplitude $\bra{\tilde\phi} \frac{b_0}{L_0}\bar{P}\ket{cV * cV}$ to calculate $\Delta B_{\phi}$ to the next-to-leading order. Such a calculation can be done again using the trick of \cite{Berkovits:2003ny}.

\section{Outlook}

We have shown that string field theory naturally tames the divergences found in conformal perturbation theory \cite{Gaberdiel:2008fn, Schmidt-Colinet:2009pga, Komargodski:2016auf}.  This is analogous to observations made by Sen in \cite{Sen:2019jpm} who showed that string field theory tames UV divergences on the string worldsheet. It would be interesting to carry our calculations to higher orders and see how far one can follow the RG flow and whether nice patterns emerge. For this probably different gauges in string field theory might be useful, such as the $\mathcal{B}_0$-gauge, possibly modified by accounting for the ${\cal L}_0 =0$ terms found in \cite{Kiermaier:2007ba} or the pseudo-$\mathcal{B}_0$ gauge introduced in \cite{Schnabl:2007az}. Similar calculations have been reported in \cite{Larocca:2017pbo}. Interestingly, our calculations seem to allow for the possibility to follow the RG flow in the opposite direction, contrary to general expectations, but in line with numerical calculations in OSFT \cite{Kudrna:2014rya}.

Finally let us mention, that our calculations are entirely feasible in {\it closed string field theory}, see \cite{Mukherji:1991tb}, where one should be able to learn about the elusive vacuum of the closed string. Our preliminary results show that the value of the closed string classical action $S$ evaluated on the solution is proportional in the leading order to the change in the central charge $\Delta c$ of the CFT where the condensation happens. Dilaton couplings seem to affect only the higher order contributions, so the result appears to be in contradiction with recent results in \cite{Erler:2022agw}. A possible resolution is that one has to incorporate the Liouville sector even for the critical string, or perhaps more likely that the higher order obstruction due to ghost dilaton will destabilize the perturbative solution.

\section*{Acknowledgments}
\noindent

We thank Mat\v{e}j Kudrna, Jakub Vo\v{s}mera, Tom\'{a}\v{s} Proch\'{a}zka and Xi Yin for useful discussions. This work has been supported by Grant Agency of the Czech Republic, under the grant EXPRO 20-25775X. We thank the Centro de Ciencias de Benasque Pedro Pascual for hospitality and providing a stimulating environment during the final stages of this work, as well as the participants of the Matrix Models and String Field Theory workshop for fruitful discussions.

\end{document}